\documentstyle[12pt]{article}
\def\a{\alpha}
\def\b{\beta}
\def\c{\chi}
\def\d{\delta}
\def\e{\epsilon}

\def\g{\gamma}

\def\p{\psi}

\def\l{\lambda}
\def\m{\mu}
\def\n{\nu}

\def\s{\sigma}

\def\vf{\varphi}

\def\be{\begin{equation}}
\def\ee{\end{equation}}
\def\arr{\begin{array}{rll}}
\def\ea{\end{array}}
\def\bea{\begin{eqnarray}}
\def\eea{\end{eqnarray}}

\begin{document}
\begin{titlepage}

\vskip 3.0cm

\begin{center}

{\Large\bf Complete Lagrangian formulation}\\

\bigskip

{\Large\bf for $N=4$ topological string}\\

\bigskip

\vskip 2cm

Stefano Bellucci\footnote{E-mail: bellucci@lnf.infn.it}
and Anton Galajinsky\footnote{On leave from Tomsk Polytechnical 
University, Tomsk, Russia\\
\phantom{XX}  
E-mail: agalajin@lnf.infn.it},
\vskip 0.5cm

{\it INFN--Laboratori Nazionali di Frascati, C.P. 13, 
00044 Frascati, Italy}

\vskip 0.2cm

\begin{abstract}
We give a Lagrangian and display all local symmetries for the
$N=4$ topological string by Berkovits, Vafa and Siegel, the latter
previously known in the superconformal gauge. Leading to a small
$N=4$ superconformal algebra and exhibiting the manifest Lorentz 
invariance the model is proposed to be a framework for restoring 
the Lorentz invariance in $N=2$ string scattering amplitudes.
\end{abstract}

\vspace{0.5cm}

\end{center}


\end{titlepage}

\noindent

The critical $N=2$ string, or a string theory exhibiting two local 
supersymmetries on the world--sheet, provides a conventional framework for a
stringy description of self--dual gauge theory and self--dual gravity
in two spatial and two temporal dimensions (or a four-dimensional
Euclidean space). Apart from a number of salient features
characterising the model (good reviews can be found 
in~\cite{markus,lechtenfeld}) like the presence of a finite number 
of physical states
in the quantum spectrum, the spectral flow phenomenon etc., 
there is a fundamental
drawback intrinsic to the theory. It lacks manifest Lorentz covariance.
Some time ago it has been realized~\cite{siegel, berkvafa,bvw} that there
exists an alternative description for the model based on a 
small $N=4$ superconformal algebra (SCA). According to the $N=4$
topological prescription, the $N=2$ SCA is to be extended by a set
of new currents which, however, prove to be functionally 
dependent on the former~\cite{siegel,bg} and do not imply any new 
information. The crucial point regarding the new setting is that
an external global automorphism group of a small $N=4$ SCA turns out to
be larger than that of the $N=2$ SCA and includes the full Lorentz group
$SO(2,2)=SU(1,1)\times SU(1,1)'$. It is interesting to mention that,
to a large extent, the situation resembles what happens for the
conventional $N=1$, $d=10$ Green--Schwarz superstring, where
extracting an irreducible subset from the original
fermionic constrains is known to be in a conflict with the manifest
Lorentz covariance. Notice further that applying the $N=4$ topological
formalism, as originally formulated in Refs.~\cite{siegel, berkvafa},
one works directly in the superconformal gauge and a gauge invariant action
necessary for a conventional path integral quantization is missing.
In view of the last obstacle, a set of reasonable
prescriptions to calculate scattering amplitudes has been proposed
in Ref.~\cite{berkvafa} and shown to reproduce the $N=2$ quantum string 
scattering. However, appealing to the topological twist, the method
of Ref.~\cite{berkvafa} does not treat all the currents on equal footing and
violates the manifest Lorentz covariance. Although the $N=4$ topological 
prescription does suggest a number of impressive simplifications
in the explicit evaluation of scattering amplitudes the main advantage 
offered by the framework, namely the manifest Lorentz covariance, seems
to be unexploited.

Recently, we constructed an action functional for the $N=4$ topological
string~\cite{bg}. The key observation was that, apart from the ordinary 
$U(1,1)$ global group which is a symmetry of the $N=2$ string
and a trivial automorphism of the $N=2$ SCA, on the space of the fields 
at hand one can realize an extra $U(1,1)$ group (sticking to the
terminology of Ref.~\cite{bvw} we call it ${U(1,1)}_{outer}$ ) 
which automatically generates the currents of a small $N=4$ SCA when 
applied to those of the $N=2$ SCA. Having recognised that, the construction
of the action amounts to the installation of the extra invariance in the
$N=2$ string action by making use of the ordinary Noether procedure.
Based on the Hamiltonian considerations we argued how many local symmetries
the $N=4$ topological string action must exhibit but failed to
give the explicit form. It is the purpose of this paper to display all
local symmetries of the model thus completing the Lagrangian formulation.
Given the latter, the standard path integral quantization 
becomes available suggesting an intriguing possibility to restore 
the manifest Lorentz covariance in the $N=2$ string scattering amplitudes.
The last point provides the main motivation for the present work.

Since the $N=4$ topological formalism relies essentially upon the 
conventional $N=2$ string we first briefly outline 
the structure of the latter. As is typical of string theories 
exhibiting an $N$--extended local supersymmetry on the world--sheet, 
the $N=2$ model describes a coupling of an $N=2$ world--sheet (on-shell) 
supergravity to matter multiplet. The latter involves a 
complex scalar $z^n$ and a complex Dirac spinor $\psi_A^n$, with the target   
index $n$ taking values $n=0,1$ in the critical dimension, while the
former is composed of a graviton ${e_n}^\a$ ($n$ stands for a flat index), 
a complex gravitino $\chi_{A \a}$, $A=1,2$, and a real vector field $A_\a$. 
The action functional reads~\cite{brink} (our $d=2$ spinor notations
are those of Ref.~\cite{bg})
\bea\label{S_0}
&& S_{N=2}=-{\textstyle{\frac {1}{2\pi}}}
\int \!\! d \tau d \s \sqrt{-g}
\{ g^{\a\b}  \partial_\a z \partial_\b {z}^{*} 
-i{\bar \psi} \g^n \partial_\a \psi {e_n}^\a 
 + i \partial_\a {\bar\psi} \g^n \psi {e_n}^\a  +
 {\bar\psi} \g^n \psi A_\a {e_n}^\a \nonumber\\[2pt]
&& \qquad \qquad +(\partial_\a z -
{\textstyle{\frac 12}} {\bar\c}_\a \psi) {\bar \psi} \g^n \g^m \c_\b 
{e_n}^\b {e_m}^\a + (\partial_\a {z}^{*} -
{\textstyle{\frac 12}} {\bar\psi} \c_\a) {\bar \c}_\b \g^n \g^m \psi 
{e_n}^\a {e_m}^\b \},
\eea
where ${*}$ stands for the complex conjugation.
Notice is that the model lacks manifest Lorentz 
covariance. This is due to the complex geometry intrinsic to the 
formalism which fails to be compatible with
the full target--space Lorentz group $Spin(2,2)=SU(1,1)\times SU(1,1)'$ 
and chooses a specific subgroup $U(1,1)\simeq  U(1)\times SU(1,1)$
as an ultimate global symmetry of the theory. 

Turning to local symmetries, with $N$ growing a world--sheet supergravity
intrinsic to a theory entails a richer family of local transformations. 
For the case at hand, apart from the ordinary diffeomorphisms, local 
Lorentz transformations and Weyl symmetry, the model reveals two new 
bosonic symmetries with real parameters $a$ and $b$
\be\label{a}
\d A_\a =\partial_\a a, \quad \d \psi=-{\textstyle{\frac {i}{2}}}
a \psi, \quad \d \c_\a  =-{\textstyle{\frac {i}{2}}}
a \c_\a ; 
\ee 
\bea\label{b}
&& \d A_\a =e^{-1} \e_{\a\b} g^{\b\g} \partial_\g b, 
\quad \d \psi=-{\textstyle{\frac {i}{2}}}
b \g_3 \psi, \quad \d \c_\a ={\textstyle{\frac {i}{2}}}
b \g_3 \c_\a,
\eea 
where $e^{-1}={(det({e_{n}}^\a))}^{-1}=\sqrt{-g}$, the super 
Weyl transformation with a complex fermionic parameter $\n_A$
\bea\label{nu}
&& \d \c_\a =-{\textstyle{\frac {i}{\sqrt{2}}}}
g_{\a\b} (\g_3 \g^n \n) {e_n}^\b, \quad
\d A_\a=-{\textstyle{\frac {i}{2\sqrt{2}}}} 
({\bar\c}_\b \g_3 \g_n \g^k \n+ {\bar\n} \g^k \g_n \g_3 \c_\b)
{e_{k}}^\b {e_\a}^n,
\eea
and an $N=2$ local world--sheet supersymmetry (the parameter is complex)
\bea\label{locsusy}
&& \d z ={\bar \e} \psi, \quad \d \c_\a =
\nabla_\a \e + {\textstyle{\frac {i}{2}}} \e A_\a
+{\textstyle{\frac {1}{8}}}(i{\bar\c}_\a \g^n \c_\g -i{\bar\c}_\g \g^n \c_\a)
(\g_n \g^m \e) {e_m}^\g,
\nonumber\\[2pt]
&& \d \p= -{\textstyle{\frac {i}{2}}}
(\partial_\g z -{\bar\c}_\g \p) (\g^n \e) {e_n}^\g, \quad 
\d {e_n}^\a={\textstyle{\frac {1}{2}}} 
(i{\bar\e} \g^k \c_\g -i {\bar\c}_\g \g^k \e)
{e_k}^\a {e_n}^\g ,
\nonumber\\[2pt]
&&  \d A_\a ={\textstyle{\frac {1}{2}}} e \e^{\b\g}
({\bar\e} \g_n \g_3 \nabla_\b \c_\g 
+\nabla_\b {\bar\c}_\g \g_n \g_3 \e){e_\a}^n
-{\textstyle{\frac {1}{4}}} e \e^{\b\g}
A_\g (i{\bar\e} \g_n \g_3 \c_\b 
\nonumber\\[2pt]
&& \quad -i{\bar\c}_\b \g_n \g_3 \e){e_\a}^n
-{\textstyle{\frac {1}{16}}} e \e^{\d\l}({\bar\c}_\l \g^k \c_\d)
(i{\bar\e} \g^m \g_k \g_n \g_3 \c_\b 
-i{\bar\c}_\b \g_3 \g_n \g_k \g^m \e)
{e_m}^\b {e_\a}^n,
\eea
where $\nabla_\a$ stands for the conventional 
world--sheet covariant derivative $\nabla_\a B_\b=\partial_\a B_\b -
\Gamma_{\a\b}^\g B_\g$. A straightforward verification of the world--sheet 
supersymmetry~(\ref{locsusy}) turns out to be pretty tedious. The hint 
is to gather the terms with a given power of $\partial z$ (or 
$\partial {\bar z}$, or both) and show that they do cancel each other.
The Fiertz identities involving the world--sheet spinors 
(recall $\g_n \g_m =-\eta_{nm}-\e_{nm} \g_3$, $\e^{nm} \g_m =\g_3 \g^n$
and $\e^{mn} \e^{kl} =-\eta^{mk}\eta^{nl}+\eta^{ml}\eta^{nk}$)
\bea\label{fiertz}
&& (\bar\p \vf) (\bar\c \g_3 \l)-(\bar\p \g_3 \vf) (\bar\c \l)=
(\bar\c \g^n \vf) (\bar\p \g_n \g_3 \l),
\nonumber\\[2pt]
&& (\bar\p \vf) (\bar\c \g_3 \l)+(\bar\p \g_3 \vf) (\bar\c \l)=
-(\bar\p \l) (\bar\c \g_3 \vf)-(\bar\p \g_3 \l) (\bar\c \vf),
\eea
prove to be crucial here. Altogether the transformations given above 
allow one to gauge away the
fields which furnish the world--sheet supergravity multiplet, while the 
variation of the action with respect to the fields made prior to the 
gauge fixing proves to yield precisely the $N=2$ superconformal currents 
as the essential constrains on the system.

Turning to the $N=4$ topological formalism, we first introduce the notation
\be
\underline{AB}=A^n \e_{nm} B^m, \qquad \underline{AB}=-{(-1)}^{\e_A \e_B}
\underline{BA},
\ee
where $\e_A$ stands for the Grassmann parity of a field $A$ and $\e_{mn}$
is the Levi-Civita antisymmetric tensor, $\e_{01}=-1$. Then    
the conventional $U(1,1)$ transformations act on the elementary field
combinations in the following way ($AB=A^n \eta_{nm} B^m$)
\vspace{0.1cm}
$$
\begin{array}{lll}
\begin{tabular}{|l|c|c|c|c|c|c|c|c|c|c|c|c|}
\hline          \vphantom{$\displaystyle\int$}
$U(1,1)$  & $\d_{\a_1}$ & $\d_{\a_2}$ & $\d_{\a_3}$ & $\d_{\a_4}$ \\
\hline          \vphantom{$\displaystyle\int$}
$\p {\vf}^{*}$ & 0 & 0 & 0 & 0\\
\hline          \vphantom{$\displaystyle\int$}
$\underline{\p\vf}$ & 0 & $-2i\a_2 \underline{\p\vf}$  & 0 & 0\\
\hline          \vphantom{$\displaystyle\int$}
$\underline{{\p}^{*} {\vf}^{*} }$ & 0 & 2i$\a_2 \underline{{\p}^{*}
{\vf}^{*}}$  & 0 & 0\\
\hline
\end{tabular}
\end{array}
$$
while the action of the ${U(1,1)}_{outer}$ group we mentioned above
reads (for more details see Ref.~\cite{bg})
\vspace{0.1cm}
$$
\begin{array}{lll}
\begin{tabular}{|l|c|c|c|c|c|c|c|c|c|c|c|c|}
\hline          \vphantom{$\displaystyle\int$}
${U(1,1)}_{outer}$  & $\d_{\a_1}$ & $\d_{\a_2}$ & $\d_{\a_3}$ & $\d_{\a_4}$ \\
\hline          \vphantom{$\displaystyle\int$}
$\p {\vf}^{*} $ & 0 & $\a_2( \underline{\p\vf}-\underline{{\p}^{*}{\vf}^{*}})$ 
& 0 & $i\a_4(\underline{\p\vf}+\underline{{\p}^{*}{\vf}^{*}})$\\
\hline          \vphantom{$\displaystyle\int$}
$\underline{\p\vf}$ & 0 & $\a_2(\p{\vf}^{*}-{\p}^{*}\vf)$ & $-2i\a_3 
\underline{\p \vf}$  & 
$-i\a_4(\p{\vf}^{*} -{\p}^{*}\vf)$\\
\hline          \vphantom{$\displaystyle\int$}
$\underline{{\p}^{*}{\vf}^{*}}$ & 0 & $-\a_2(\p{\vf}^{*}-{\p}^{*}\vf)$ &
 2i$\a_3 \underline{{\p}^{*}{\vf}^{*}}$  & 
$-i\a_4(\p{\vf}^{*} -{\p}^{*}\vf)$\\
\hline
\end{tabular}
\end{array}
$$
\vspace{0.1cm}
The action functional which exhibits both $U(1,1)$ and ${U(1,1)}_{outer}$
global invariance requires the introduction of some extra gauge 
fields on the world--sheet of the string~\cite{bg}. These turn out 
to complement
an $N=2$, $d=2$ supergravity multiplet to that of an 
$N=4$, $d=2$ supergravity and include two real vector fields   
$B_\a$ and $C_\a$ and a complex spinor $\m_{A \a}$ (see for example
Ref.~\cite{gates}). The new fields transform nontrivially under the action 
of ${U(1,1)}_{outer}$ group~\cite{bg} 
\vspace{0.1cm}
$$
\begin{array}{lll}\label{19}
\begin{tabular}{|l|c|c|c|c|c|c|c|c|c|c|c|c|}
\hline          \vphantom{$\displaystyle\int$}
${U(1,1)}_{outer}$  & $\d_{\a_1}$ & $\d_{\a_2}$ & $\d_{\a_3}$ & $\d_{\a_4}$ \\
\hline          \vphantom{$\displaystyle\int$}
$A_\a$ & 0 & $-4\a_2 B_\a$ 
& 0 & $-4\a_4 C_\a$\\
\hline          \vphantom{$\displaystyle\int$}
$B_\a$ & 0 & $-\a_2 A_\a$ & $-2\a_3 C_\a$  & 0\\
\hline          \vphantom{$\displaystyle\int$}
$C_\a$ & 0 & 0 &
 $2\a_3 B_\a$  & 
$-\a_4 A_\a$\\
\hline          \vphantom{$\displaystyle\int$}
$\m_\a$ & 0 & $-\a_2 {(\c -{\c}^{*})}_\a$ 
& $2i\a_3 \m_\a$ & $-i\a_4 {(\c -{\c}^{*})}_\a $\\
\hline          \vphantom{$\displaystyle\int$}
$\c_\a$ & 0 &  $-\a_2 {(\m -{\m}^{*})}_\a$    
& 0  &   $i\a_4 {(\m +{\m}^{*})}_\a $ \\
\hline          
\end{tabular}
\end{array}
$$
\vspace{0.1cm}
the bosons furnishing a triplet representation with respect to
${SU(1,1)}_{outer}$ subgroup while the fermions forming the 
doublet representation. Conformed to the conventional spinor 
notation the action functional of Ref.~\cite{bg} reads
\bea\label{S}
&& S_{N=4}=S_{N=2}-{\textstyle{\frac {1}{2\pi}}}
\int \!\! d \tau d \s \sqrt{-g}
\{ -\underline{\p C\g^n \p} (B_\a +iC_\a){e_n}^\a 
+ \underline{{\p}^{*} C \g^n {\p}^{*}} (B_\a -iC_\a){e_n}^\a
\nonumber\\[2pt]
&& \qquad +\underline{\partial_\a z (\p C \g^n \g^m \m_\b)}{e_n}^\b {e_m}^\a  
+\underline{\partial_\a z^{*} (\m_\b^{*} C \g^m \g^n {\p}^{*})}{e_n}^\b 
{e_m}^\a  
\nonumber\\[2pt]
&& \qquad 
-{\textstyle{\frac {1}{2}}}({\bar\c}_\a \g^m \g^n \m_\b)
\underline{(\p C\g_m \g^k \p )} {e_n}^\a {e_k}^\b 
+{\textstyle{\frac {1}{2}}}({\bar\m}_\b \g^n \g^m \c_\a)
\underline{({\p}^{*} C\g^k \g_m {\p}^{*} )} {e_n}^\a {e_k}^\b
\nonumber\\[2pt] 
&& 
\qquad 
-{\textstyle{\frac {1}{8}}}(\m_\a C \g^m \g^n \underline{\p)(\p} C
\g_m \g^k {(\c+{\c}^{*})}_\b )({e_n}^\b {e_k}^\a+{e_n}^\a {e_k}^\b)
\nonumber\\[2pt]
&& \qquad 
+{\textstyle{\frac {1}{8}}}({(\c+{\c}^{*})}_\b C \g^k \g^m 
\underline{{\p}^{*})({\p}^{*}} C
\g^n \g_m \m_\a^{*})({e_n}^\b {e_k}^\a+{e_n}^\a {e_k}^\b)
\nonumber\\[2pt]
&& \qquad 
+{\textstyle{\frac {1}{2}}}({\bar\m}_\b \g^m \g^n \m_\a )(\bar\p \g_m \g^k
\p) {e_n}^\b {e_k}^\a \}.
\eea
Here $C$ stands for the charge conjugation matrix which in the
representation chosen just coincides with $\g_0$ (see also the
Appendix of Ref.~\cite{bg}).

We now turn to discuss local symmetries of the full action. The first
important point to check is that adding the new fields to the original
$N=2$ theory one does not spoil the symmetries~(\ref{a})-(\ref{locsusy}). 
For the bosonic $a$-- and $b$--transformations this turns out to be the case
provided the following transformation laws for the newly introduced fields
\bea
\d \m_\a = {\textstyle{\frac {i}{2}}} a \m_\a, 
\quad \d {(B +i C)}_\g=i a {(B +i C)}_\g-{\textstyle{\frac {i}{4}}} a
g^{\a\b} (\m_\a C \g_n \c_\b) {e_\g}^n,
\eea
and
\bea
\d \m_\a =-{\textstyle{\frac {i}{2}}} b \g_3  \m_\a, 
\quad \d {(B +i C)}_\g=i b e^{-1} \e_{\g\d} g^{\d\a}
{(B +i C)}_\a+{\textstyle{\frac {i}{4}}} b
g^{\a\b} (\m_\a C \g_n \g_3 \c_\b) {e_\g}^n.
\eea
For the super Weyl transformations~(\ref{nu}) to hold in the extended $N=4$ action
the complex gauge field ${(B+iC)}_\a$ is to be transformed according to
the rule
\bea
\d {(B +i C)}_\g={\textstyle{\frac {i}{4\sqrt{2}}}}
(\m_\a C \g_n \g^k \g_3 (\n+\n^{*})) {e_k}^\a {e_\g}^n. 
\eea 
Finally, the $N=2$ local supersymmetry requires
\bea
&& \d \m_\b=i\e{(B+iC)}_\b +{\textstyle{\frac {i}{8}}}
(\g^n \g^m \m_\b) ({(\c-\c^{*})}_\g C \g_n \e) {e_m}^\g
+{\textstyle{\frac {i}{8}}}
(\g^n \g^m \m_\g) ({(\c-\c^{*})}_\b C \g_n \e) {e_m}^\g
\nonumber\\[2pt]
&& \quad \quad +{\textstyle{\frac {i}{4}}} (\g^n \g^m \m_\g)
(\bar\e \g_n \c_\b)  {e_m}^\g,\nonumber\\[2pt]
&& \d {(B+iC)}_\a = {\textstyle{\frac {1}{4}}}
(\bar \e \g_n \g^k \g^m \nabla_\l \m_\b) {e_k}^\b {e_m}^\l {e_\a}^n
-{\textstyle{\frac {1}{4}}}
(\partial_\l \bar \e \g_n \g^k \g^m \m_\b) {e_k}^\b {e_m}^\l {e_\a}^n
\nonumber\\[2pt]
&& \quad \quad
+{\textstyle{\frac {i}{4}}}
(\bar \e \g_n \g^k \g^m \c_\g-{\bar \c}_\g \g^m \g^k \g_n \e)
{(B+iC)}_\l {e_k}^\g {e_m}^\l {e_\a}^n \nonumber\\[2pt]
&& \quad \quad 
-{\textstyle{\frac {i}{4}}}
(\bar \e \g_n \g^k \g^m \c_\g+{\bar \c}_\g \g^m \g^k \g_n \e)
{(B+iC)}_\l {e_k}^\l {e_m}^\g {e_\a}^n
\nonumber\\[2pt]
&& \quad \quad +{\textstyle{\frac {i}{16}}}
(\bar \e \g_k \g^n \g^m \c_\g+{\bar \c}_\g \g^m \g^n \g_k \e)
(\m_\l C \g_n {(\c+\c^{*})}_\b)g^{\l\b} {e_m}^\g {e_\a}^k
\nonumber\\[2pt]
&& \quad \quad  
+{\textstyle{\frac {i}{16}}}
(\bar \e \g^k \g^n \g^l \c_\l-{\bar \c}_\l \g^l \g^n \g^k \e)
(\m_\g C \g^a \g_n \g^b {(\c-\c^{*})}_\d) {e_k}^\g {e_l}^\d {e_b}^\l
{e_\a}^a\nonumber\\[2pt]
&& \quad \quad
+{\textstyle{\frac {1}{4}}}{e_\a}^n \e_{nm} \d(\m_\l
C \g^m {(\c+\c^{*})}_\b g^{\l\b}),
\eea 
for the extra fields and moreover, the variations of $\p^n$ and
$A_\a$ are to include the new contributions
\bea
&& \d_{add} A_\a= {\textstyle{\frac {i}{16}}} (\bar\e \g^k \g^n \g^p \c_\g-
{\bar\c}_\g \g^p \g^n \g^k \e)({\bar\m}_\l \g_l \g_n \g^m \m_\d
+{\bar\m}_\d \g^m \g_n \g_l \m_\l) {e_k}^\l {e_p}^\d {e_m}^\g {e_\a}^l,
\nonumber\\[2pt]
&& \d_{add} \p^n=-{\textstyle{\frac {i}{4}}} \e^{nm} (\g^p \e)
(\m_\g^{*} C \g_p \g^k \p_m^{*}) {e_k}^\g, 
\eea
in order to provide a symmetry of the extended formalism.
Notice that the last two lines involve explicitly the Levi--Civita
tensor, thus providing a nontrivial interplay between $AB$-- and
$\underline{AB}$--terms.

As we have mentioned earlier, on the world--sheet the theory is described
by an $N=4$, $d=2$ supergravity multiplet. The latter proves to entail a 
richer family of local symmetries as compared to those provided by
an $N=2$, $d=2$ supergravity multiplet
underlying the conventional $N=2$ string. A complete Hamiltonian 
analysis for the model at hand has been performed in Ref~\cite{bg}.
The conclusion based on the counting of unspecified Lagrange multipliers 
corresponding to primary first class constraints in the question 
was that the theory must exhibit two more bosonic
symmetries with complex parameters, these corresponding to the (complex)
gauge field ${(B+iC)}_\a$, and two more fermionic symmetries with
complex spinor parameters, these coming along with $\m_{A\a}$ and being the
partners for the super Weyl symmetry and the local supersymmetry. 
Making use of the identities~(\ref{fiertz}) one indeed finds two new bosonic
transformations which resemble very much the $a$-- and $b$--symmetries  
(in contrast to the transformations~(\ref{a}) and ~(\ref{b}) the parameters 
$c$ and $f$ are complex)
\bea
&& \d \p^n =c \e^{nm} \p_m^{*}, \quad \d \c_\a =-c \m_\a,
\quad \d \m_\a =-{\bar c} \c_\a,\nonumber\\[2pt]
&& \d A_\a = -2c {(B+iC)}_\a-2 {\bar c}{(B-iC)}_\a+
{\textstyle{\frac {1}{2}}} c (\m_\b C \g_n \c_\d) g^{\b\d} {e_\a}^n
+{\textstyle{\frac {1}{2}}} {\bar c} (\c_\b^{*} C \g_n \m_\d^{*}) g^{\b\d} 
{e_\a}^n,\nonumber\\[2pt]
&& \d {(B+iC)}_\a =-i \partial_\a {\bar c} -{\bar c} A_\a
+{\textstyle{\frac {1}{4}}} {\bar c} ({\bar\m}_\b \g_n \m_\d -
{\bar\c}_\b \g_n \c_\d) g^{\b\d} {e_\a}^n, 
\eea
and
\bea
&& \d \p^n =f \e^{nm} \g_3 \p_m^{*}, \quad \d \c_\a =f \g_3 \m_\a,
\quad \d \m_\a ={\bar f} \g_3 \c_\a,\nonumber\\[2pt]
&& \d {(B+iC)}_\a =-i {e}^{-1} \e_{\a\b} g^{\b\g}
\partial_\g {\bar f} -{\bar f} {e}^{-1} \e_{\a\b} g^{\b\g} A_\g 
-{\textstyle{\frac {1}{4}}} {\bar f}
({\bar\m}_\b \g_n \g_3 \m_\g -
{\bar\c}_\b \g_n \g_3 \c_\g) g^{\b\g} {e_\a}^n, \nonumber\\[2pt]
&& \d A_\a = -2f {e}^{-1} \e_{\a\b} g^{\b\g} {(B+iC)}_\g
-2 {\bar f} {e}^{-1} \e_{\a\b} g^{\b\g} {(B-iC)}_\g+
{\textstyle{\frac {1}{2}}} f (\c_\g C \g_n \g_3 \m_\b) g^{\b\g} {e_\a}^n
\nonumber\\[2pt]
&& \quad \quad 
+{\textstyle{\frac {1}{2}}} {\bar f} (\m_\b^{*} C \g_n \g_3 \c_\g^{*})
g^{\b\g} {e_\a}^n.
\eea
In its turn the super Weyl transformation prompts the form 
of a partner realized on the extra fields
(the parameter $\l_A$ is a complex $d=2$ spinor) 
\bea
\d \m_\a =g_{\a\b} (\g_3 \g^k \l) {e_k}^\b, \quad
\d {(B+iC)}_\a ={\textstyle{\frac {1}{4}}}(\d \m_\g C \g_n 
{(\c+{\c}^{*})}_\b) g^{\g\b} {e_\a}^n. 
\eea
Finally, with the extensive use of the identities~(\ref{fiertz}) one can verify
the invariance of the action under a novel fermionic symmetry
\bea
&& \d z^n =-\e^{nm} (\p_m^{*} C \e^{*}), \quad  
\d {e_n}^\a=-{\textstyle{\frac {i}{4}}}( \bar\e \g_n \g^m \g^k \m_\b
-{\bar\m}_\b \g^k \g^m \g_n \e) {e_m}^\b {e_k}^\a,
\nonumber\\[2pt]
&& \d \p^n=-{\textstyle{\frac {i}{2}}} \e^{nm} \partial_\g z_m^{*}
(\g^k \e^{*}) {e_k}^\g +{\textstyle{\frac {i}{4}}} \e^{nm}
(\g^p \p_m^{*}) (\bar\e \g^k \g_p \c_\g){e_k}^\g+
{\textstyle{\frac {i}{4}}} (\g^p \p) (\bar\e \g^k \g_p \m_\g) {e_k}^\g,
\nonumber\\[2pt]
&& \d \m_\b =-\nabla_\b \e + {\textstyle{\frac {i}{2}}} \e A_\b
+{\textstyle{\frac {i}{4}}} (\g^n \g^m \m_\g)(\bar\e \g_n \m_\b -
{\bar\m}_\b \g_n \e){e_m}^\g-{\textstyle{\frac {i}{8}}} 
(\g^n \g^m \e)({\bar\c}_\b \g_n \c_\g+{\bar\c}_\g \g_n \c_\b){e_m}^\g,
\nonumber\\[2pt]
&& 
 \d \c_\b =i\e{(B-iC)}_\b  
+{\textstyle{\frac {i}{4}}} (\g^n \g^m \c_\g)(\bar\e \g_n \m_\b -
{\bar\m}_\b \g_n \e){e_m}^\g-{\textstyle{\frac {i}{8}}} 
(\g^n \g^m \e)({\bar\m}_\b \g_n {(\c+\c^{*})}_\g
\nonumber\\[2pt]
&& \qquad \qquad +{\bar\m}_\g \g_n {(\c+\c^{*})}_\b){e_m}^\g,
\nonumber\\[2pt]
&& \d A_\a =-{\textstyle{\frac {1}{4}}} 
(\nabla_\g {\bar\m}_\b \g_k \g^n \g^m \e
+\bar\e \g^m \g^n \g_k \nabla_\g \m_\b) {e_n}^\b {e_m}^\g {e_\a}^k
-{\textstyle{\frac {1}{2}}} {e_\a}^n \e_{nm} \d ({\bar\c}_\b \g^m \c_\g 
g^{\g\b})  
\nonumber\\[2pt]
&& +{\textstyle{\frac {i}{4}}}(\bar\e \g^k \g^n \g_m \m_\b
-{\bar\m}_\b \g_m \g^n \g^k \e) A_\g {e_n}^\b {e_k}^\g {e_\a}^m
+{\textstyle{\frac {1}{4}}} (\partial_\g \bar\e \g_k \g^n \g^m \m_\b
+{\bar\m}_\b \g^m \g^n \g_k \partial_\g \e) {e_n}^\b {e_m}^\g {e_\a}^k
\nonumber\\[2pt]
&& +{\textstyle{\frac {i}{16}}} (\bar\e \g^k \g^n \g^p \m_\b -
{\bar\m}_\b \g^p \g^n \g^k \e) ({\bar\m}_\l \g_l \g_n \g^m \m_\g
+{\bar\m}_\g \g^m \g_n \g_l \m_\l){e_m}^\b {e_k}^\l {e_p}^\g
{e_\a}^l,
\nonumber\\[2pt] 
&& \d {(B+iC)}_\a ={\textstyle{\frac {1}{4}}}({\bar\c}_\b \g^k \g^n \g_m
\partial_\g \e){e_n}^\b {e_k}^\g {e_\a}^m -
{\textstyle{\frac {1}{4}}}(\nabla_\g {\bar\c}_\b \g^k \g^n \g_m 
\e){e_n}^\b {e_k}^\g {e_\a}^m 
\nonumber\\[2pt]
&&+{\textstyle{\frac {i}{4}}}(\bar\e \g^k \g^n \g_m \m_\b -{\bar\m}_\b \g_m
\g^n \g^k \e) {(B+iC)}_\g {e_n}^\b {e_k}^\g {e_\a}^m
\nonumber\\[2pt]
&& +{\textstyle{\frac {1}{4}}} {e_\a}^n \e_{nm} \d(\m_\g C \g^m
{(\c+\c^{*})}_\b g^{\g\b})
+ {\textstyle{\frac {i}{16}}}(\bar\e \g^k \g^n \g^m \m_\g
-{\bar\m}_\g \g^m \g^n \g^k \e)\times 
\nonumber\\[2pt]
&& (\m_\d C \g_l \g_n \g^p {(\c-\c^{*})}_\b)
{e_m}^\b {e_k}^\d {e_p}^\g {e_\a}^l,
\eea
this complementing the transformations~(\ref{locsusy}) to the full $N=4$ 
local supersymmetry. It should be stressed that a straightforward 
verification of the extra fermionic symmetry turns out to be
pretty tedious. Getting rid of $\gamma$--matrices like we did in the 
previous work~\cite{bg} both in the action and in the transformation laws
simplifies the calculation considerably since does not appeal to any Fiertz
identities. The form of the transformations, however, look more compact in 
the conventional spinor notations to which we stick here.

To summarise, in this paper we gave the action (conformed to
the conventional spinor notation) and established
all local symmetries for the $N=4$ topological string. A natural next step
which, to a great extent, motivated the present investigation, 
is to construct scattering amplitudes proceeding from
the conventional path integral quantization. The reducibility of the current
forming an $N=4$ SCA will presumably cause certain complications of the
conventional method. However, since all the currents are scalars one 
can expect to reveal a finite stage of reducibility
for the system and hence finitely many ghosts involved. 

\vspace{0.4cm}
\noindent
{\bf Acknowledgements}\\[-4pt]

\noindent
We thank Nathan Berkovits for useful discussions. The work was supported 
by INTAS grant No 00-00254 and by the Iniziativa Specifica MI12 of the 
Commissione IV of INFN.

\end{document}